\def\DraftSize{
 \typeout{Miserly format}
 \documentstyle{article}
 \textwidth 172mm
 \textheight 230mm
 \topmargin -40pt
 \oddsidemargin -20pt
 \renewcommand{\baselinestretch}{0.95}
 }

\def\FullSize{
 \typeout{Generous format}
 \documentstyle[12pt]{article}
 \textwidth17cm
 \textheight22cm
 \oddsidemargin0em
 \topmargin-2ex
 \parindent=3em
}

\DraftSize

\makeatletter
\@addtoreset{equation}{section}

\makeatother

\def\beq{\begin{equation}}
\def\eeq{\end{equation}}
\def\bea{\begin{eqnarray}}
\def\nn{\nonumber \\ }
\def\eea{\end{eqnarray}}
\def\ds{\displaystyle}
\def\nsz{\normalsize}

\def\ni{\noindent}

\def\req#1{(\ref{#1})}

\def\rep{{\rm Re}\ }

\def\scapt#1{\caption{\protect\small #1}}
\def\barc{\begin{array}{c}}
\def\ear{\end{array}}

\renewcommand{\baselinestretch}{1.4}

\title{
Influence of a magnetic fluxon on the
vacuum energy of quantum fields confined by a bag.
}
\author{
\nsz S. Leseduarte$^a$\thanks{E-mail: lese@ecm.ub.es} \
and August Romeo$^{b \, c}$\thanks{E-mails: august@ceab.es, romeo@ieec.fcr.es}, \\
\nsz $^a$ \it Dept ECM and IFAE, Faculty of Physics, University of
Barcelona,  \\
\nsz\it Diagonal 647, 08028 Barcelona  \\
\nsz $^b$ \it Blanes Centre for Advanced Studies (CEAB), CSIC, \\
\nsz\it Cam{\'\i} de Santa B{\`a}rbara, 17300 Blanes  \\
\nsz $^c$ \it Institut d'Estudis Espacials de Catalunya (IEEC), \\
\nsz\it Edifici Nexus, c. Gran Capit{\`a} 2-4, 08034 Barcelona
}

\date{}

\begin{document}

\maketitle
\begin{abstract}
We study the simultaneous influence of
boundary conditions and external fields on quantum fluctuations by
considering
vacuum zero-point energies for quantum fields in
the presence of a magnetic fluxon confined by
a bag,
circular and spherical for bosons and
circular for fermions.
The Casimir effect is calculated in a 
generalized cut-off regularization
after applying zeta-function techniques to eigenmode
sums and using recent techniques about Bessel zeta functions at
negative arguments. 

\end{abstract}

\section{Introduction}

Aharonov-Bohm \cite{AB} settings may be regarded as one of the
possible ways in which an external field modifies some observables
of a given quantum system. In this specific phenomenon, the presence
of an infinitely thin tube of magnetic flux alters the energy spectrum
and brings about a modification of the vacuum energy, giving rise
to a form of Casimir effect.
Initially, Aharonov-Bohm fields acted on free particles\footnote{For
Casimir interactions between {\it two} solenoids see \cite{Du}}.
The purpose of the present paper is to study the influence of the
same type of fields on systems which are already constrained by boundary
conditions (b.c.), working out the combined net effect of both on what
would otherwise be a free system.
The relevance of the Aharonov-Bohm scenario to some cosmic strings models
including particles in the gravitational field of a spinning source
is discussed in \cite{JdSG,dSG}. A similar mathematical procedure is also applied
to the description of Dirac fermions on black-hole backgrounds as shown in
\cite{Kabat}.

The quantum mechanical problem of a scalar particle inside
a circular Aharonov-Bohm quantum billiard \cite{Ber}-\cite{Ste}
(of radius $a$) 
bears a great resemblance, from the mathematical point of view, to the ones
we set out to consider.
In our case we have a classical magnetic fluxon which is coupled to
a quantum field. We take this object to be an idealization of a vortex
with a radially symmetric distribution of magnetic field in the limit
where its characteristic thickness is vanishingly small.
With this model in mind one has the physical basis to fix
the boundary conditions at the origin to be imposed on the modes
of the matter field, (a detailed account of this kind of analysis
is given in \cite{AMW}).
We start with
a complex, Klein-Gordon, massless field (the additional
analytic effort which takes the treatment of a massive field by zeta
function techniques is explained in \cite{MBEEKKSL}).
We
call $\phi$ the space-dependent part of the eigenmodes, which
satisfies the equation (in units such that $\hbar=c=1$)
\beq
\left( -i\vec{\nabla} -e\vec{A} \right)^2 \phi = \omega^2\phi,
\label{Schr}
\eeq
where the vector potential $\vec{A}$ is given by
\beq
e \vec{A}(\vec{r})= {\alpha \over r} \hat{e}_{\varphi},
\hspace{1cm}
\alpha={e\Phi\over 2\pi}.
\label{Aalpha}
\eeq
$\alpha$ is called {\it reduced flux}, being $\Phi$ the flux of the
magnetic field.
Since a billiard is a domain with perfectly reflecting walls, and we
imagine an infinitely thin solenoid at the origin
---reduced, in $D=2$, to an unreachable point---
the b.c. are $\phi=0$ at $r=0$ and $r=a$.

Zero-point energies emerge from mode-sums of the type
$\ds{1 \over 2}\sum_n \omega_n$, and give rise
to the Casimir effect \cite{Cas}-\cite{AW}.
Since the summation extends over all the $\omega_n$'s in the set of
eigenmodes, such quantities
usually diverge and need some regularization to make sense
of them.
To this end,
we introduce spectral zeta functions as mere
auxiliary tools, which will be denoted by
\beq
\zeta_{{\cal M}}(s)=\sum_n \omega_n^{-s}, \hspace{1cm}
\zeta_{{\cal M}\over \mu}(s)=
\sum_n \left( \omega_n\over \mu \right)^{-s}.    \label{formaldef}
\eeq
$\mu$ is an arbitrary scale with mass dimensions, used to work
with dimensionless objects. This is a regularization of analytical
nature (see also \cite{SaSt}); other examples in this same category
are the techniques in refs. \cite{McShRe} and
refs. \cite{Speer}, \cite{Francia}.
When we discuss the results, we shall
comment on their physical significance from the perspective of cut-off
regularization. In this sense, our standpoint in the present case
is that the zeta function is
a purely mathematical object which affords a convenient method for the calculation
of observables inasmuch as it may be connected with other, more physical,
regularizations. Let us assume that we are using a general cut-off
regularization
for the vacuum energy which is given by
\beq
E_{reg}=\frac{1}{2} \sum_{n} \omega_n \,
g \left( \frac{\omega_n}{\Lambda} \right) ,
\label{cutoff}
\eeq
where $g$ is a well-behaved function which satisfies asymptotic
expansions near the origin and at infinity of the following kind:
\beq
g(t) \sim 1+\sum_{k=1}^{\infty} a_k t^k\,\,\,(t\rightarrow 0) \hspace{1cm}
g(t) \sim \frac{1}{t^{M_s}} \sum_{k=0}^{\infty } b_k t^{-k}
\,\,\,(t\rightarrow \infty). \label{asymptotic}
\eeq
If we restrict this analysis for the sake of simplicity to
a $(2+1)-D$ case,
then $g$ should be such that $M_s>3$.
This conditions guarantee that the Mellin transform of $g$ has no poles in
the strip $0<{\Re}z<M_s$.
The
$\zeta_{\cal M}$-function,
defined in (\ref{formaldef}), has
its rightmost pole at $s=2$.
In fact, the requirement which we demand on the
asymptotic behaviour of $g$ might
have been stated with greater generality. Essentially,
$g$ should go to 1 for small
values of its argument, and should go to zero for large values fast
enough so that the following reasoning makes sense. Now
we explain presently how the connection
between the $\zeta_{\cal M} $-function (\ref{formaldef})
and expression (\ref{cutoff}) may be accomplished
(for other discussions on how the results
from different regulators may be related, see \cite{CoVaZe,BeSa}).

The Parseval formula for Mellin transforms allows us to write
\beq
E_{reg}(\Lambda)=\frac{1}{2} \frac{1}{2\pi i} \int_{r-i\infty}^{r+i\infty}
dz\,\Lambda^{1+z}\,{\rm M}\left[ g,z+1\right] \zeta_{\cal M} (z),
\label{integrep}
\eeq
where $r$ is any real number such that $2<r<M_s-1$. The structure of the
poles of the
integrand in expression (\ref{integrep}) is closely
related to the asymptotic
expansion of $E_{reg}(\Lambda)$ for large values of the cut-off $\Lambda$.
In the case at issue the relevant poles (the rightmost ones)
are found at $z=2$, $z=1$
and $z=-1$. The first one is a simple pole due to the
divergence
of $\zeta_{\cal M} $ at $z=2$. This induces
the strongest divergence in $E_{reg}$ (the
one that is related to the volume (surface)
and which in our two dimensional case goes
as $\Lambda ^3$).
The next divergence arises from the pole of $\zeta_{\cal M} $
at $z=1$
and goes as $\Lambda ^2$. The last divergence stems from a pole
at $z=-1$.
This pole is in general a double one because
both ${\rm M}\left[ g,z+1\right] $ and $\zeta_{\cal M} $
have poles at that point. This fact means that by properly applying the Cauchy
theorem, the integrand at $z=-1$ determines the finite part of $E_{reg}$,
and also a divergent piece which goes as the logarithm of
$\Lambda $.
Let us give a more detailed account of these remarks. From the
hypothesis we have stated, we may write
\bea
{\rm M}\left[ g,\,z\right] & =& \frac{1}{z} + {\rm Fin}\left[ {\rm M}
\left[ g,\,z=0\right] \right] + {\cal O}(z) \nn
\zeta(z)&=&\frac{{\rm Res}\left[ \zeta ,\,2\right] }{z-2}
+{\cal O}( (z-2)^0 ) \nn
\zeta(z)&=&\frac{{\rm Res}\left[ \zeta ,\,1\right] }{z-1}
+{\cal O}( (z-1)^0 ) \nn
\zeta(z)&=&\frac{{\rm Res}\left[ \zeta ,\,-1\right] }{z+1}
+{\cal O}( (z+1)^0 ) ,
\label{pols}
\eea
where the symbol ${\rm Fin}$ means the extraction of the finite coefficient
from the Laurent expansion of a function.
Now we may use these expressions in \req{integrep} to get that, apart
from terms which go to zero for large values of the cut-off $\Lambda $,
the regularized expression for the vacuum energy is
\bea
E_{reg}(\Lambda ) &=& \frac{1}{2}\left( \Lambda^3 {\rm M}\left[ g,\,3\right] \,
{\rm Res}\left[ \zeta ,\,2\right] + \Lambda^2 {\rm M}\left[ g,\,2\right] \,
{\rm Res}\left[ \zeta ,\,1\right] \right. \nn
&&+\left. \ln\Lambda{\rm Res}\left[ \zeta,\,-1\right] +{\rm Fin}\left[ {\rm M}
\left[ g\right] ,\,z=0\right] \,\,{\rm Res}\left[ \zeta,\,-1\right] +
{\rm Fin}\left[ \zeta,\,z=-1\right] \right) .\label{lamexpansion}
\eea

The conclusion
of this analysis is that to prove that the divergent
terms in $E_{reg}(\Lambda )$
are independent of the magnetic flux, it suffices to show that the residues of
$\zeta_{\cal M}$ at $z=2$,
$z=1$ and $z=-1$ do not depend on this physical parameter.
Let us put
it another way, if we label the different heat-kernel
coefficients
$B_0,\,B_{\frac{1}{2}},\,B_1,\,B_{\frac{3}{2}} \ldots $,
this independence boils down to saying
that $B_0$, $B_1$ and $B_{\frac{3}{2}}$ 
do not depend on the magnetic flux. As for the finite
part of expression (\ref{cutoff}) and the quantity
${\rm Fin}\left[ \zeta,\,z=-1\right] $,
their
relationship is quite direct,
as it appears explicitly in expression \req{lamexpansion}.

We shall systematically throw away such divergences which do not 
depend on $\alpha$. These divergences would be relevant in a study
about the bag dynamics (see \cite{BVW,MBEEKKSL}), 
that is, concerned with situations where
the bag walls are liable to deformation. In our case, we take the
bag to be a perfectly rigid object. 

In the free case (i.e. without flux) the eigenfrequencies under
our conditions  are zeros of $J_{\nu}$  Bessel functions
with integer indices $\nu $ coming from angular momentum.
The solutions for nonzero $\alpha$
have been found in \cite{Ber,Ste}, and basically correspond to an index
shift with respect to the free case
$|l| \to |l-\alpha|$.
Since in
both cases the
eigenmodes are zeros of the same type of functions,
we shall introduce the following `partial-wave' zeta functions
for fixed values of $\nu$:
\beq
\zeta_{\nu}^{}(s)=
\sum_{n=1}^{\infty} j_{\nu, n}^{-s} \, ,  \
\mbox{for $\rep s > 1$},
\label{zetanuSe}
\eeq
where $j_{\nu, n}$ denotes the $n$th nonvanishing zero of $J_{\nu}$
(see also \cite{ELR,LR};
discrete versions of the Bessel problem, their solutions and
associated zeta functions have also been studied in
\cite{Kvit}).

When considering the whole problem in a $D$-dimensional space,
one must take into account
the degeneracy $d(D,l)$ of each angular mode in $D$ dimensions.
Therefore, we define the `complete' spherical zeta function
\begin{equation}
\begin{array}{lllll}
\ds\zeta_{\cal M}^{}(s)&=&
\ds a^s \sum_{l=l_{\rm min}}^{\infty} d(D,l)
\sum_{n=1}^{\infty} j_{ \nu(D,l), n}^{-s}
&=&\ds a^s \sum_{l=l_{\rm min}}^{\infty} d(D,l) \,
\zeta_{\nu(D,l)}^{}(s) , \\
\end{array}
\label{defzdDs}
\end{equation}
$l_{\rm min}$ is the minimum value (if any) of $l$.
In the free case
$\nu(D,l)=l+D/2-1$ and
the general form of $d(D,l)$ (see e.g.    \cite{Vil}) is
$\ds d(D,l)=(2l+D-2) { (l+D-3)! \over l! (D-2)! }$ , but this
will change when a flux is present.

Following the programme we have just put forward,
the partial wave zeta function for scalars is obtained in sect. 2.
  From this starting point,
we construct the complete zeta functions for $D=2$ and $D=3$
complex scalar fields in sects. 3 and 4, respectively,
finding their analytic continuations to $s=-1$.
Numerical results for the zero-point energy are then discussed.
Afterwards, in sect. 5 we study the Dirac field in $D=2$, and sect. 6
is devoted to the conclusions.

\section{`Partial-wave' zeta function}

Computing the Casimir energy through the calculation of the
complete zeta function  requires
the knowledge of the Bessel zeta functions \req{zetanuSe} at $s=-1$,
while
the complex domain where \req{zetanuSe} holds is bounded by $\rep s =1$.
This is a serious difficulty, but we know that
$\zeta_{\nu}^{}(s)$ admits
an analytic continuation to other values of $s$.
Moreover,
in refs. \cite{ELR} and \cite{LR} we showed how to obtain an
integral representation of this continuation
valid for $-1 < {\rm Re}\  s < 0$, which reads
\begin{equation}
\zeta_{\nu}^{}(s)={s \over \pi} \sin{\pi s \over 2}
\int_0^{\infty} dx \, x^{-s-1}
\ln\left[ \sqrt{2\pi x} \, e^{-x} I_{\nu}(x) \right] ,  \  \
\mbox{for $-1 < {\rm Re}\  s < 0$}.
\label{zetanuID}
\end{equation}
Whenever $\nu\neq 0$ we can
work out (\ref{zetanuID}) by the method explained in
\cite{RU,LRap} (see also \cite{Bar} and \cite{BoKi}),
arriving at
\begin{equation}
\begin{array}{lll}
\ds\zeta_{\nu}^{}(s)&=&
\ds{1 \over 4}\sigma^{}_1 \nu^{-s} \\
&&\ds+\nu^{-s} {s \over \pi} \sin{\pi s \over 2}\left[
\sigma^{}_2
\left\{
\ds{1 \over 2s} B\left( {s+1 \over 2}, -{s \over 2} \right)
+ 2^{s-1} B\left( {s+1 \over 2}, -s \right) \right.\right. \\
&&\hspace{9em}\ds\left.+ 2^{s-1} B\left( {s+3 \over 2}, -s \right)
\right\} \nu \\
&&\hspace{6em}\ds\left. +{\cal S}^{}_N(s, \nu)
+{1 \over 2} \rho^{}
B\left( {s+1 \over 2}, -{s \over 2} \right){1 \over \nu}
+\overline{\cal J}^{}_1(s) {1 \over \nu}
+\sum_{n=2}^{N} {\cal J}^{}_n(s) {1 \over \nu^n} \right] .
\end{array}
\label{myzetanus}
\end{equation}
with
\begin{equation}
\sigma^{}_1=-1, \hspace{1cm}
\sigma^{}_2= 1, \hspace{1cm}
\rho^{}={1 \over 8}.
\label{sigmasandrho}
\end{equation}
In addition
\begin{equation}
\begin{array}{lll}
\ds {\cal S}_N^{}(s, \nu)&\equiv&
\ds\int_0^{\infty} dx \, x^{-s-1} \, \left\{
\ln\left[
{\cal L}^{}(\nu, x)
\right]
-\sum_{n=1}^N { {\cal U}^{}_n(t(x)) \over \nu^n }
\right\} , \\
\ds{\cal L}^{}(\nu, x)&=&\ds
\sqrt{2\pi\nu}(1+x^2)^{1/4} e^{-\nu\eta(x)} I_{\nu}(\nu x) ,
\hspace{1cm}
\eta(x)= \sqrt{1+x^2}+\ln{x \over 1+\sqrt{1+x^2}} ,
\label{defSJ}
\end{array}
\end{equation}
\begin{equation}
\begin{array}{lll}
\ds{\cal U}^{}_1(t)&=&\ds
{t\over 8} - {{5\,{t^3}}\over {24}}, \\
\ds{\cal U}^{}_2(t)&=&\ds
{{{t^2}}\over {16}} - {{3\,{t^4}}\over 8} +
 {{5\,{t^6}}\over {16}}, \\
\ds{\cal U}^{}_3(t)&=&\ds
{{25\,{t^3}}\over {384}} - {{531\,{t^5}}\over {640}} +
     {{221\,{t^7}}\over {128}} - {{1105\,{t^9}}\over {1152}}, \\
\ds{\cal U}^{}_4(t)&=&\ds
     {{13\,{t^4}}\over {128}} -
     {{71\,{t^6}}\over {32}} +
     {{531\,{t^8}}\over {64}} -
     {{339\,{t^{10}}}\over {32}} +
     {{565\,{t^{12}}}\over {128}} , \\
&\vdots&
\end{array}
\label{U14}
\end{equation}
the key point being that, this way, $S^{}_N(s, \nu)$
is a {\it finite} integral at $s=-1$.
Further,
\beq
\begin{array}{lll}
\ds\overline{\cal J}^{}_1(s)&=&\ds-{5 \over 48}
B\left( {s+3 \over 2}, -{s \over 2} \right) \\
\ds {\cal J}^{}_n(s)&=&\ds \int_0^{\infty} dx \,
x^{-s-1} \, {\cal U}^{}_n(t(x)), \hspace{2em}
t(x)={1 \over \sqrt{1+x^2}},
\end{array}
\label{JsfromUs}
\eeq
Thus, the expressions for the ${\cal J}^{}_n(s)$'s
are easily obtained from the ${\cal U}^{}_n(t)$'s in
(\ref{U14}). In fact, since
\beq
\int_0^{\infty} dx \, x^{-s-1} \, [t(x)]^m
={1 \over 2}B\left( {s+m \over 2}, -{s \over 2} \right) ,
\label{inttm}
\eeq
the result of the $x$-integration is like replacing
\beq
\begin{array}{rcl}
\ds{\cal U}^{}_n(t)&\to&
\ds{\cal J}^{}_n(s) \\
t^m&\to&\ds{1 \over 2}B\left( {s+m \over 2}, -{s \over 2} \right) .
\end{array}
\eeq

Expression \req{myzetanus} is not valid for $\nu=0$, since it was
obtained from a rescaling $x \to \nu x$ and subsequent
application of uniform asymptotic expansions in $\nu x$.
Moreover,
numerically speaking it is little convenient if $\nu$ is very small.
An alternative representation valid in these conditions is  needed.
Starting from \req{zetanuID},
we subtract and add the asymptotic behaviour of the
integrand, which gives rise to a logarithmic divergence on integration.
When doing so, we shall write the large-$x$ expansion of $\ln[ \dots ]$
as follows:
\beq
\ln\left[
\sqrt{2\pi x}
\, e^{-x} I_{\nu}(x) \right]
=-{4\nu^2-1 \over 8x}+{\cal O}\left( 1 \over x^2 \right)
=-{4\nu^2-1 \over 8\sqrt{x^2+1} }+{\cal O}\left( 1 \over x^2+1 \right) .
\eeq
Thus, the piece we separate can be integrated with the help of
\req{inttm} ($m=1$ case) and we are left with
\beq
\begin{array}{lll}
\zeta_{\nu}^{}(s)&=&\ds
{s \over \pi} \sin{\pi s \over 2} \left[
{\cal R}_{\nu}(s) -{4\nu^2-1 \over 16}
B\left( {s+1 \over 2}, -{s \over 2} \right) \right] ,\\
{\cal R}_{\nu}(s)&=&\ds
\int_0^{\infty} dx \, x^{-s-1} \left\{
\ln\left[
\sqrt{2\pi x}
\, e^{-x} I_{\nu}(x)
\right]
+{4\nu^2-1 \over 8\sqrt{x^2+1} } \right\} .
\end{array}
\eeq
Since the above integral is now finite at $s=-1$ we can Laurent-expand
without problems around $s=-1$, arriving at
\beq
\zeta_{\nu}^{}(s)=
{1-4\nu^2 \over 8\pi}{1 \over s+1}
+{1-4\nu^2 \over 8\pi}(-1+\ln 2)
+{1 \over\pi}{\cal R}_{\nu}(-1)
+{\cal O}(s+1) .
\label{zetanuID0le}
\eeq
In particular, for $\nu=0$, ${\cal R}_{0}(-1)=-0.00723$ and
\beq
\zeta_{0}^{}(s)=
{1 \over 8\pi}{1 \over s+1} -0.01451
+{\cal O}(s+1) .
\eeq
Note also the vanishing of the $s=-1$ pole when $\nu=\pm 1/2$  already
explained in ref. \cite{LR}.

\section{$D=2$ bosons}

\subsection{`Complete' zeta function}

Next, we go on to the two-dimensional problem.
Following ref.\cite{Ste}, one realizes that the eigenmode sum for
this case gives rise to
the following complete spectral zeta function
\beq
\zeta_{\cal M}(s ; \alpha)= a^s
\sum_{l=-\infty}^{\infty} \zeta_{ |l-\alpha| }^{}(s),
\eeq
Since this function has the properties
\beq
\begin{array}{lll}
\zeta_{\cal M}(s ; \alpha+k)&=&\zeta_{\cal M}(s ; \alpha) , \
k\in{\bf Z}, \\
\zeta_{\cal M}(s; -\alpha)&=&\zeta_{\cal M}(s ; \alpha) ,
\end{array}
\label{perbarzb}
\eeq
it is enough to study it for $0\leq \alpha \leq 1/2$.
Introducing
\beq
\overline{\zeta_{\cal M}}(s ; \beta)\equiv a^s
\sum_{l= 0}^{\infty} \zeta_{ l+\beta  }^{}(s) ,
\label{defbarz}
\eeq
we can write
\beq
\begin{array}{lll}
\ds\zeta_{\cal M}(s ; \alpha)
&=&\ds\overline{\zeta_{\cal M}}(s ; \alpha)
+ \overline{\zeta_{\cal M}}(s ; 1-\alpha) \\
&=&\ds a^s\zeta_{|\alpha|}^{}(s)
+\overline{\zeta_{\cal M}}(s ; 1+\alpha)
+ \overline{\zeta_{\cal M}}(s ; 1-\alpha) .
\end{array}
\label{czf3p}
\eeq
Next we insert expression \req{myzetanus} into \req{defbarz} and,
realizing that
$\ds
\sum_{l=0}^{\infty}(l+\beta)^{-s}=\zeta_H(s,\beta),
$
where $\zeta_H$ stands for the Hurwitz zeta function, we find
\begin{equation}
\begin{array}{lll}
\ds\overline{\zeta_{\cal M}}(s;\beta)&=&\ds
{1 \over 4} \sigma^{}_1 a^{s} \zeta_H(s,\beta) \\
&&\ds+a^{s} {s \over \pi} \sin{\pi s \over 2}\left[
\sigma^{}_2
\left\{
\ds{1 \over 2s} B\left( {s+1 \over 2}, -{s \over 2} \right)
+ 2^{s-1} B\left( {s+1 \over 2}, -s \right) \right.\right. \\
&&\ds\hspace{9em}\left. + 2^{s-1} B\left( {s+3 \over 2}, -s \right)
\right\} \zeta_H(s-1, \beta) \\
&&\hspace{6em}\ds +\sum_{l=0}^{\infty }{\cal S}^{}_N(s, l+\beta)
(l+\beta)^{-s}\\
&&\hspace{6em}\ds +{1 \over 2} \rho^{}
B\left( {s+1 \over 2}, -{s \over 2} \right) \zeta_H(s+1,\beta) \\
&&\hspace{6em}\ds\left.
+\overline{\cal J}^{}_1(s) \zeta_H(s+1,\beta)
+\sum_{n=2}^{N} {\cal J}^{}_n(s) \zeta_H(s+n,\beta)
\right] ,
\end{array}
\label{barz}
\end{equation}
with the values of $\sigma_1$, $\sigma_2$ and $\rho$ in
\req{sigmasandrho}.
Taking $N=4$ and Laurent-expanding near $s=-1$, this may be written
\beq
\begin{array}{ll}
\ds\overline{\zeta_{\cal M}}(s;\beta)={1 \over a}&\ds\left[
-{1 \over 4}\zeta_H(-1,\beta) \right. \\
&\ds+{1 \over \pi} \left\{
{1 \over 4}\zeta_H(-2,\beta)
-{5 \over 24}\zeta_H(0, \beta)
-{229 \over 40320}\zeta_H(2,\beta)
+{35 \over 65536}\zeta_H(3,\beta) \right. \\
&\hspace{2em}\ds+\sum_{l=0}^{\infty} S^{}_4(-1,l+\beta)(l+\beta) \\
&\hspace{2em}\ds
+\left(
-{\pi\over 256}
-{1\over 2}\zeta_H(-2,\beta)
+{1 \over 8}\zeta_H(0, \beta )
\right)
\left( {1 \over s+1}+\ln a -1 \right) \\
&\hspace{2em}\ds-{\pi \over 64}+{\ln 2 \over 16}-\beta{\ln 2 \over 8}
+{\pi\psi(\beta) \over 256}
-\left( 1+{1\over 2}\ln 2 \right)\zeta_H(-2, \beta) \\
&\hspace{2em}\ds\left.\left.
-{1 \over 2}\zeta_H'(-2, \beta)+{1 \over 8}\zeta_H'(0,\beta)
\right\}
+{\cal O}(s+1)
\right] .
\end{array}
\label{barzle}
\eeq
Concerning the pole at $s=-1$ of the complete zeta function, by
\req{czf3p}, \req{zetanuID0le} and \req{barzle}, and noticing that
$\zeta_H(-2,1+\alpha)+\zeta_H(-2,1-\alpha)=-\alpha^2$, one arrives at
\beq
\ds\zeta_{\cal M}(s ; \alpha)=
{1 \over a}\left[
-{1 \over 128}{1 \over s+1}+{\cal O}((s+1)^0)
\right],
\eeq
i.e. the residue is independent of $\alpha$.
The reader may check by using the method explained in ref.\cite{BoKi}
that this independence applies not only to $B_{\frac{3}{2}}$, but also
to $B_0$ and $B_{\frac{1}{2}}$.
As we have explained in the introduction,
this property allows us to state that in cut-off regularization the
dependence of the vacuum energy on the magnetic flux does not appear in
the divergent terms. The dependence of the vacuum energy
on the magnetic flux is completely contained in the finite part of
$\zeta_{\cal M}$.

Since we plan to use the same three formulas for calculating the
finite parts, it will be necessary to
obtain
$\zeta_H'( 0, \beta)$ and $\zeta_H'( -2,\beta)$
around $\beta=1$. The first is known (see e.g. \cite{BMP}) and
amounts to
\beq
\zeta_H'(0,\beta)=\ln\Gamma(\beta)-{1 \over 2}\ln(2 \pi) ,
\label{zhp0}
\eeq
while the second is calculated by numerical evaluation of
 $\zeta_H'( -n,\beta)$ from an integral representation of
the derivative of $\zeta_H$ valid for negative first arguments.

\subsection{Numerical results}
We start by the $l=0$ partial wave zeta-functions obtained from
\req{zetanuID0le}. Since we are supposing $\alpha \geq 0$,
the results will be denoted by
\beq
a^s\zeta_{ \alpha }^{}(s)=
{1 \over a}\left[
r_{\alpha} \left( {1 \over s+1} +\ln a \right)
+p_{\alpha} +{\cal O}(s+1) \right] ,
\hspace{1cm}
r_{\alpha}={1-4\alpha^2 \over 8\pi},
\eeq
where
the finite parts $p_{\alpha}$ are
listed in the second column of table \ref{tbpw}.
\begin{table}[htbp]
\begin{center}
\begin{tabular}{|l|r|r|r|r|}
\hline
$\alpha$&$p_{\alpha}$&$\bar{p}_{1+\alpha}$&$\bar{p}_{1-\alpha}$&
$q_{\alpha}$ \\ \hline
$0  $&$-0.01451$&$+0.01174$&$+0.01174$&$+0.00899$ \\
$0.1$&$-0.05971$&$+0.04062$&$-0.01172$&$-0.03081$ \\
$0.2$&$-0.10771$&$+0.07491$&$-0.02987$&$-0.06266$ \\
$0.3$&$-0.15778$&$+0.11462$&$-0.04285$&$-0.08601$ \\
$0.4$&$-0.20932$&$+0.15968$&$-0.05095$&$-0.10060$ \\
${1 \over 2}$&$-{\pi \over 12}=-0.26180$&$+0.21001$&$-0.05471$&$-0.10650 $\\
\hline
\end{tabular}
\end{center}
\scapt{Finite parts at $s=-1$ of the involved zeta funcions (for $a=1$).
Column 2: $l=0$ partial wave zeta function $\zeta_{\alpha}(s)$
Columns 3 and 4:
$\overline{\overline{\zeta_{\cal M}}}(s ; \beta )$ with
$\beta=1+\alpha$ and $\beta=1-\alpha$.
Column 5: finite part of the
complete zeta function $\zeta_{\cal M}(s ; \alpha)$.}
\label{tbpw}
\end{table}
The pole absence for $\alpha=1/2$ may be regarded as a
consequence of the fact that $J_{1/2}(x) \propto \sin x$, and therefore
$\zeta_{1/2}(x)=\pi^{-s}\zeta_R(s)$ ($\zeta_R$ meaning the Riemann
zeta function), which is finite at $s=-1$ because $\zeta_R(-1)=-1/12$.
Next, we find $\overline{\zeta_{\cal M}}(s; \beta)$ from \req{barzle}
for the corresponding $\beta=1\pm\alpha$'s.
We shall employ the notation
\beq
\overline{\zeta_{\cal M}}(s; \beta)=
{1 \over a}\left[
\bar{r}_{\beta} \left( {1 \over s+1} +\ln a \right)
+\bar{p}_{\beta} \right] +{\cal O}(s+1) .
\eeq
According to \req{barzle}
\beq
\bar{r}_{\beta}=
{1 \over \pi} \left[
-{\pi\over 256}
-{1\over 2}\zeta_H(-2, \beta )
+{1 \over 8}\zeta_H(0, \beta)
\right]
\label{barrbeta}
\eeq
(note that in terms of Bernoulli polynomials
$\zeta_H(-n,x)=-{1 \over n+1}B_{n+1}(x)$).
As for $\bar{p}_{\beta}$, we list some of its values
in columns 3 and 4 of table \ref{tbpw}.
Using now \req{czf3p} and the above results we get
\beq
\zeta_{\cal M}(s;\alpha)={1 \over a}
\left[
-{1 \over 128}\left( {1 \over s+1} +\ln a \right) + q_{\alpha}
\right] +{\cal O}(s+1) ,
\eeq
The already remarked $\alpha$-indepedence of the resdiue is
exhibited by the fact that
$\ds
r_{\alpha}+\bar{r}_{1+\alpha}+\bar{r}_{1-\alpha}= -{1 \over 128}
$.
Values of
$q_{\alpha}=p_{\alpha}+\bar{p}_{1+\alpha}+\bar{p}_{1-\alpha}$
for different $\alpha$'s between 0 and $1/2$ are given in
the fifth column of table \ref{tbpw} (see also Fig. 1).
Now it would be incorrect to say that the dependence of $E_{reg}(\Lambda)$
on $\alpha $ is exactly given by ${1 \over 2a} \, q_\alpha $.
We should take into account
a factor 2 which stems from the complex nature of the scalar field. In
other words, the dependence of $E_{reg}(\Lambda)$ on $\alpha $ is given
by ${1 \over a} \,q_\alpha$, apart from terms which vanish when
$\Lambda$ goes to infinity.

\section{$D=3$ bosons}

Eq. \req{Schr} is again considered, but now in $D=3$ and with a
magnetic flux line diametrically threading a sphere of radius $a$.
We make such a gauge choice that the vector potential in spherical
coordinates reads
\beq
e {\vec{A}}(\vec{r})=
{\alpha \over r \sin\theta } \hat{e}_{\varphi} .
\eeq

The spectrum and eigenfuctions for the associated quantum-mechanical
problem have been written down by the authors in ref.\cite{ELR}.
The proof that one must impose regularity at the origin
may be carried out as described in ref.\cite{AMW}.
After studying their associated degeneracies,
we are able to write the complete zeta function
as follows
\beq
\zeta_{\cal M}(s ; \alpha)=
a^s \sum_{p=0}^{\infty} \sum_{m=-\infty }^{\infty} \zeta_{|m-\alpha |+p+1/2}(s),
\label{genuine3D}
\eeq
again, it is apparent that
\beq
\zeta_{\cal M}(s ; \alpha + k)=
\zeta_{\cal M}(s ; \alpha)     \label{period3D1}
\eeq
for any integer $k$, and
\beq
\zeta_{\cal M}(s ; 1-\alpha)=
\zeta_{\cal M}(s ; \alpha) .    \label{period3D2}
\eeq
It is now an immediate result that
\beq
\zeta_{\cal M}(s ; \alpha)=
\zeta_{\cal M}(s ; -\alpha)     \label{period3D}
\eeq
 From \req{period3D1} and \req{period3D2} we may also restrict
our study to the domain $0 \leq \alpha \leq \frac{1}{2}$,
this is a property which we proceed to take advantge of in the sequel.
Under this restriction we may give an alternative representaion for
\req{genuine3D}:
\beq
\zeta_{\cal M}(s ; \alpha)=
a^s \sum_{l=-\infty}^{\infty} |l| \, \zeta_{ |l-\alpha+1/2| }^{}(s).
\eeq
In terms of
\beq
\widetilde{\zeta_{\cal M}}(s ; \beta ) \equiv
a^s\sum_{l=0}^{\infty} l \, \zeta_{ l+\beta }^{}(s)
\eeq
and of the  $\overline{\zeta_{\cal M}}$ function defined in \req{defbarz},
$\zeta_{\cal M}(s ; \alpha)$ reads
\beq
\zeta_{\cal M}(s ; \alpha)=
\overline{\zeta_{\cal M}}\left( s ; {1 \over 2}+\alpha \right)
+\widetilde{\zeta_{\cal M}}\left( s ; {1 \over 2}+\alpha \right)
+\widetilde{\zeta_{\cal M}}\left( s ; {1 \over 2}-\alpha \right) .
\label{barandtildes}
\eeq
Next, let's consider the relation between them and the new zeta function
\beq
\overline{\overline{\zeta_{\cal M}}}(s ; \beta) \equiv
a^s\sum_{l=0}^{\infty} (l+\beta) \zeta_{ l+\beta }^{}(s).
\eeq
This has the advantage that
$\overline{\overline{\zeta_{\cal M}}}(s ; \beta )$
can be immediately found from known material. The case without
magnetic flux has been already studied for $D=3$ in \cite{RU,LRap}.
Since $d(3,l)=2l+1=2\nu(3,l)$, formula \req{defzdDs} is now
rewritten as
$
\ds\zeta_{\cal M}(s; \alpha=0 )=
2 a^s \sum_{l=0}^{\infty} \nu(3,l) \zeta_{\nu(3,l)}(s)
$.
Therefore, the expression for
$\overline{\overline{\zeta_{\cal M}}}(s ; \beta )$ is
the one for the $\ds\zeta_{\cal M}(s)$ in those works but for
the simple replacement
\beq
\overline{\overline{\zeta_{\cal M}}}(s ; \beta )=
{1 \over 2} \, \zeta_{\cal M}(s; \alpha=0)
\left\{ \nu(l)=(l+1/2) \longrightarrow (l+\beta ) \right\}  .
\label{replbarbar}
\eeq
Thus, for $N=4$ subtractions we find
\beq
\begin{array}{ll}
\ds\overline{\overline{\zeta_{\cal M}}}(s ; \beta )=
{1 \over a}&\ds\left[
-{1 \over 256}\zeta_H(0, \beta)
+{1 \over 4\pi}\zeta_H(-3, \beta)
-{1 \over 4}\zeta_H(-2, \beta)
-{5 \over 24 \pi}\zeta_H(-1, \beta)
+{35 \over 65536}\zeta_H(2, \beta) \right. \\
&\ds+{1 \over \pi}\left\{
\left(
-{229 \over 40320}
-{1\over 2}\zeta_H(-3,\beta )
+{1\over 8}\zeta_H(-1,\beta )
\right)
\left( {1 \over s+1} +\ln a -1 \right) \right.\\
&\ds\hspace{2em}+\sum_{l=0}^{\infty}
{\cal S}^{}_4 (-1, l+\beta) (l+\beta)^2 \\
&\ds\hspace{2em}+
{293 \over 24192}
-{229 \over 40320}( \ln 2 - \psi(\beta) ) \\
&\ds\hspace{2em}
+\left( -1 -{1 \over 2}\ln 2 \right) \zeta_H(-3,\beta)
-{1 \over 2}\zeta_H'(-3,\beta) \\
&\ds\hspace{2em}\left.\left.+{1 \over 8}\ln 2 \, \zeta_H(-1,\beta)
+{1 \over 8} \zeta_H'(-1,\beta) \right\}
+{\cal O}(s+1)
\right] .
\end{array}
\label{BBZsb}
\eeq
As a result of the previous definitions,
\beq
\widetilde{\zeta_{\cal M}}(s ; \beta ) =
\overline{\overline{\zeta_{\cal M}}}(s ; \beta )
-\beta \overline{\zeta_{\cal M}}(s ; \beta ) .
\eeq
Hence, the complete zeta function \req{barandtildes}
is conveniently put in the way
\beq
\zeta_{\cal M}(s ; \alpha)= \ds
\left( {1 \over 2} - \alpha \right) \left[
\overline{\zeta_{\cal M}}\left( s ; {1 \over 2}+\alpha \right)
-\overline{\zeta_{\cal M}}\left( s ; {1 \over 2}-\alpha \right)
\right]
+\overline{\overline{\zeta_{\cal M}}}\left( s ; {1 \over 2}+\alpha \right)
+\overline{\overline{\zeta_{\cal M}}} \left( s ; {1 \over 2}-\alpha \right) ,
\label{complbars}
\eeq
and the necessary knowledge about the objects on the r.h.s. is available.
We have already found the residue of
$\overline{\zeta_{\cal M}}(s; \beta )$ at $s=-1$, namely
\beq
\mbox{Res}\left[ \overline{\zeta_{\cal M}}(s; \beta );
s=-1 \right]=
{1 \over a} \bar{r}_{\beta} ,
\eeq
where $\bar{r}_{\beta}$ is the one in \req{barrbeta}.
Similarly, from \req{BBZsb}
\beq
\mbox{Res}\left[ \overline{\overline{\zeta_{\cal M}}}(s; \beta);
s=-1 \right]=
{1 \over a\pi} \left[
-{229 \over 40320}
-{1\over 2}\zeta_H(-3,\beta )
+{1\over 8}\zeta_H(-1,\beta )
\right] .
\eeq
(At this point, one can check that
$$\ds
\mbox{Res}\left[ \overline{\overline{\zeta_{\cal M}}}(s; \beta= 1/2);
s=-1 \right] =
{1 \over a} {1 \over 315 \pi} =
\mbox{Res}\left[ {1 \over 2}\zeta_{\cal M}(s; \alpha=0); s=-1 \right] ,
$$
as should be).
Then, \req{complbars} yields
\beq
\mbox{Res}\left[ \zeta_{\cal M}(s; \alpha); s=-1 \right]=
{1 \over a\pi}\left[
{2 \over 315}
-{1 \over 6}\alpha(1-\alpha^2)\left( 1-{\alpha \over 2} \right)
\right] .
\label{resal}
\eeq
We recall that this is valid for $0 \leq \alpha \leq \frac{1}{2}$,
and one has to make use of \req{period3D1} and \req{period3D2} to
extend it to any value, in particular, when we extend expression
\req{resal} to any real $\alpha$, we have a non-analytic function of
$\alpha$. The theory cannot be renormalized.
The situation is different if we
give up the idea of a purely
confining enclosure and allow the presence of external modes, but
satisfying the same b.c. as the internal ones.
Parallelling the
steps in ref.\cite{LRap} for the $\alpha=0$ case, we construct the
$\alpha$-dependent complete zeta function for these external Dirichlet
modes ---say $\zeta_{{\cal M} \, {\rm ext}}^{}(s; \alpha)$.
With respect to the internal case, we have the following
modifications:
\beq
{\cal L}^{}(\nu, x) \to
\sqrt{2\nu\over \pi}(1+x^2)^{1/4} e^{\nu\eta(x)} K_{\nu}(\nu x),
\eeq
and the $\nu$-series undergoes a $\nu$-parity change,
which brings about the transformations
\beq
\begin{array}{lll}
\sigma_2 &\to& -\sigma_2,  \\
\rho &\to& -\rho,  \\
{\cal U}_n(t) &\to& (-1)^n {\cal U}_n(t), \\
\overline{\cal J}_1(s), {\cal J}_n(s) &\to&
-\overline{\cal J}_1(s),(-1)^n {\cal J}_n(s) .
\end{array}
\eeq
The external $\zeta $-function ensuing 
from these transformations takes into account
an overall subtraction from the Minkowsky space (for instance, the residue
of the rightmost pole is negative).
 From the construction of this external
$\zeta $-function, all the terms contributing to the $s=-1$ pole
---including a piece proportional to ${\cal J}_3(s)$ ---
reverse their sign
with respect to their internal counterparts and
\beq
\mbox{Res}
\left[ \zeta_{{\cal M} \, {\rm ext}}^{}(s; \alpha); s=-1 \right]=
-\mbox{Res}\left[ \zeta_{\cal M}^{}(s; \alpha); s=-1 \right] ,
\eeq
as a result of which  the net zeta function
$\ds \zeta_{\cal M}^{}(s; \alpha)
+\zeta_{{\cal M} \, {\rm ext}}^{}(s; \alpha)$
is finite at $s=-1$ regardless of the $\alpha$ value.
The same cancellation applies for the resudues at $s=1$ and $s=3$.
Such a cancellation is typical of odd $D$'s, and does not
happen in $D=2$ because the residue receives then a contribution from
${\cal J}_2(s)$, which maintains its sign.
To be brief, we only have to worry about the residue at $s=2$.
It is quite immediate that it is $\alpha $ independent.
The conclusion is that for a $3-D$ Klein-Gordon field defined in
both the exterior and the interior region, the whole dependence of
the vacuum energy on the $\alpha $ parameter is contained in the 
finite part of $\ds \zeta_{\cal M}^{}(s; \alpha)
+\zeta_{{\cal M} \, {\rm ext}}^{}(s; \alpha)$.

The $\alpha$-dependences of the residue $r_{\alpha}$
and of the finite part $p_{\alpha}$
of $\zeta_{\cal M}(s; \alpha)$ at $s=-1$ are depicted in Figs. 2a and 2b
for the internal modes only.
The residue is simply formula \req{resal}, while the
finite parts have been obtained through numerical evaluation of
\req{complbars} by the methods described in refs. \cite{RU,LRap}.
Fig. 2c shows the inclusion of the external modes, and the net
dependence on $\alpha$ of the vacuum energy.
Though we do not pay too much attention to the absolute figures, but
only to the dependence on $\alpha$ (in other words, to the
derivative of the finite part with respect to $\alpha $), it is
worthwhile noting that the value at $\alpha = 0$ furnishes us with
an opportunity to verify our results.
We have obtained that the value of the graph at $\alpha =0$ is
$\frac{1}{a} 0.005634... = 2 \cdot {1 \over a} 0.002817...$
i.e. twice the figure found in \cite{BeMi} for an ordinary free field,
as had to be expected.

\section{$D=2$ fermions}
For $D=2$ massless Dirac particles under the influence of the
same magnetic field as in sect. 1 and 3, the Dirac equation reads
\beq
(i \not\!{\partial} + e \not\!\!{A}) \Psi = 0,
\eeq
with
$\not\!{v} \equiv \gamma^{\mu} v_{\mu}$
($\gamma^0=\sigma^3$, $\gamma^1=i\sigma^2$, $\gamma^2=-i\sigma^1$),
$A^0(\vec{r})=0$ and $\vec{A}(\vec{r})$ as in \req{Aalpha},
(for previous works where $\zeta $-function techniques are studied
in fermionic systems see \cite{DEE,CK,ABDK}).

The boundary conditions that we choose on the bag, given by
the circle $r=a$, are those of
the M.I.T. bag model
\[ -i \not\!{n} \Psi = \Psi \]
where $n$ stands for the normal vector.

It has been remarked in  refs.\cite{dSG,JdSG} that in this problem
it would be too restrictive to impose regularity at the
origin for the modes. If one imposes regularity the result is that
the domain of the operator is not dense and, consequently, one
loses self-adjointness.
Making $\Psi( \vec x, t)= \psi( \vec x ) \, e^{-i Et}$,
let us note the space-dependent part of a particular mode by
\[ 
\psi( \vec x(r, \varphi) ) =
\left( \begin{array}{c} \chi^1(r) \\ \chi^2(r) \,e^{i\varphi} \end{array}
\right) e^{im\varphi} .
\]
It is easily seen that if one demands regularity for the modes
characterized by $m=- \left[ \alpha \right] -1$, then one is left only with
the trivial solution $\Psi =0$. As we advanced in the introduction,
we have followed the analysis which was set forth in \cite{AMW}; the
outcome is that for the particular value of $m=-\left[ \alpha \right] -1$, one
should choose the solution with regular $\chi^1$ for positive $\alpha $,
and the one with regular $\chi^2$ when $\alpha$ is negative. In any case
this amounts to picking an element
among a family of possible
self-adjoint extensions for the Hamiltonian under the b.c. in question.
Then, the whole set of Hamiltonian eigenfrequencies consists of:
\begin{enumerate}
\item the $k$'s satisfying
$f_{\nu}(ka) \equiv J_{\nu}^2(ka)-J_{\nu+1}^2(ka) = 0$ with
$\nu=\{ \alpha \}-1$  if $\alpha > 0$, and with $\nu= -\{ \alpha \}$ otherwise.
\item\begin{enumerate}
\item the $k$'s satisfying $f_{l+\{ \alpha \}}(ka)=0$, for $l=0,1,2,\dots $
\item the $k$'s satisfying $f_{l+1-\{ \alpha \}}(ka)=0$, for $l=0,1,2,\dots $
\end{enumerate}
\end{enumerate}
where $\{ \alpha \}$ denotes the fractional part of $\alpha$.

It is then adequate to define
\beq
\zeta_{\nu}^{\rm f}(s)=
\sum_{n=1}^{\infty} {\lambda}_{\nu, n}^{-s} \, ,  \
\mbox{for $\rep s > 1$},
\label{zetanuSef}
\eeq
where ${\lambda}_{\nu, n}$ means the $n$th nonvanishing zero of
$f_{\nu}(\lambda)$.
Now we may write down the fermionic zeta function as
\beq
\zeta_{\cal M}^{\rm f}(s) \equiv \zeta_{1\,{\cal M}}^{\rm f}(s) +
\zeta_{2\,{\cal M}}^{\rm f}(s) ,
\label{zeta1zeta2}
\eeq
where
\bea
\zeta_{1\,{\cal M}}^{\rm f}(s) & \equiv & \theta (\alpha )
\zeta_{ \{ \alpha \} -1}^{\rm f} (s) +\theta (-\alpha )
\zeta_{ -\{ \alpha \} }^{\rm f} (s) \label{zeta1izeta2} , \nn
\zeta_{2\,{\cal M}}^{\rm f}(s) & \equiv &
\sum_{n=0}^{\infty } \zeta_{ \{ \alpha \}
+n }^{\rm f}(s) + \sum_{n=0}^{\infty } \zeta_{1- \{ \alpha \} +n }^{\rm f}(s) .
\eea

It follows that the Casimir energy will have to fulfil the equalities
\beq
\begin{array}{c}
E_C( -\alpha )= E_C(\alpha), \\
\begin{array}{llll}
E_C(\alpha +k)&=&E_C(\alpha) ,&\mbox{for} \ \alpha > 0, k=1,2,3,
\dots  \\
E_C(\alpha -k)&=&E_C(\alpha) ,&\mbox{for} \ \alpha < 0, k=1,2,3, \dots
\end{array}
\end{array}
\label{periodferm}
\eeq
(compare with relations \req{perbarzb}).
Since we have now this sort of periodicity when shifting $\alpha$ by
integer values, it suffices for our study to take $0 \leq \alpha < 1$.

With the help of the auxiliary object
\beq
\overline{\zeta^{\rm f}_{\cal M}}(s ; \beta)\equiv a^s
\sum_{l= 0}^{\infty} \zeta^{\rm f}_{ l+\beta  }(s) ,
\label{defbarzf}
\eeq
(analogous to \req{defbarz} for bosons)
we are able to express the
complete zeta function as
\beq
\begin{array}{lll}
\ds\zeta^{\rm f}_{\cal M}(s ; \alpha)
&=&\ds a^s \zeta^{\rm f}_{\alpha-1}(s)
+\overline{\zeta^{\rm f}_{\cal M}}(s ; \alpha)
+ \overline{\zeta^{\rm f}_{\cal M}}(s ; 1-\alpha) \\
&=&\ds a^s [ \zeta^{\rm f}_{\alpha-1}(s)
            +\zeta^{\rm f}_{\alpha}(s) ]
+\overline{\zeta^{\rm f}_{\cal M}}(s ; 1+\alpha)
+ \overline{\zeta^{\rm f}_{\cal M}}(s ; 1-\alpha)
\label{czf3pf}
\end{array}
\eeq
(For the numerical methods to be applied below,
the second form proves to be more suitable around $\alpha=0$).

Starting from the partial wave zeta function \req{zetanuSef},
we make use of the technique described in \cite{ELR,LR} and find an
analytic continuation to the domain $-1 < {\rm Re}\  s < 0$
given by the integral representation
\begin{equation}
\zeta_{\nu}^{\rm f}(s)={s \over \pi} \sin{\pi s \over 2}
\int_0^{\infty} dx \, x^{-s-1}
\ln\left\{ \pi x \, e^{-2x} [ I_{\nu}^2(x)+I_{\nu+1}^2(x) ] \right\} , \  \
\mbox{for $-1 < {\rm Re}\  s < 0$}.
\label{zetanuIDf}
\end{equation}
For $\nu \neq 0$,
and by a subtraction method similar to the one applied in the bosonic
case, we obtain the more convenient form
\begin{equation}
\begin{array}{lll}
\ds\zeta_{\nu}^{\rm f}(s)&=&\ds
\nu^{-s} {s \over \pi} \sin{\pi s \over 2}\left[
\left\{
\ds{1 \over s} B\left( {s+1 \over 2}, -{s \over 2} \right)
+ 2^{s  } B\left( {s+1 \over 2}, -s \right)
+ 2^{s  } B\left( {s+3 \over 2}, -s \right)
\right\} \nu \right. \\
&&\hspace{6em}\ds +{\cal S}^{\rm f}_N(s, \nu)  \\
&&\hspace{6em}\ds \left. +{1 \over 2s} B\left( {s+1 \over 2}, -{s \over 2} \right)
-{1 \over 8}
B\left( {s+1 \over 2}, -{s \over 2} \right){1 \over \nu}
+\overline{\cal J}^{\rm f}_1(s) {1 \over \nu}
+\sum_{n=2}^{N} {\cal J}^{\rm f}_n(s) {1 \over \nu^n} \right] ,
\end{array}
\label{myzetanusf}
\end{equation}
with
\begin{equation}
\begin{array}{lll}
\ds {\cal S}_N^{\rm f}(s, \nu)&\equiv&
\ds\int_0^{\infty} dx \, x^{-s-1} \, \left\{
\ln\left[
{\cal L}^{\rm f}(\nu, x)
\right]
-\sum_{n=0}^N { {\cal U}^{\rm f}_n(t(x)) \over \nu^n }
\right\} , \\
\ds{\cal L}^{\rm f}(\nu, x)&=&\ds
[ \sqrt{2\pi\nu}(1+x^2)^{1/4} e^{-\nu\eta(x)} ]^2
{1 \over 2} [I_{\nu}^2(\nu x)+I_{\nu+1}^2(\nu x)],
\label{defSJf}
\end{array}
\end{equation}
where
\begin{equation}
\begin{array}{lll}
\ds{\cal U}^{\rm f}_0(t)&=&\ds\ln(1-t), \\
\ds{\cal U}^{\rm f}_1(t)&=&\ds
-{t\over 4} + {t^3\over 12}, \\
\ds{\cal U}^{\rm f}_2(t)&=&\ds
  {t^8\over 8} + {t^4\over 8}
- {t^5\over 8} - {t^6\over 8} , \\
\ds{\cal U}^{\rm f}_3(t)&=&\ds
 {5 t^3 \over 192} +{t^4 \over 8} +{9 t^5\over 320}
-{t^6\over 2} -{23 t^7\over 64} +{3 t^8\over 8}
-{179 t^9\over 576}, \\
\ds{\cal U}^{\rm f}_4(t)&=&\ds
     {t^4\over 32} +
     {17t^5\over 128} -
     {t^6\over 8} -
     {{165\,{t^7}}\over {128}} -
     {{37\,{t^8}}\over {64}} +
     {{327\,{t^9}}\over {128}} +
     {{57\,{t^{10}}}\over {32}} -
     {{179\,{t^{11}}}\over {128}} -
     {{71\,{t^{12}}}\over {64}} , \\
&\vdots&
\end{array}
\label{U14f}
\end{equation}
and
\beq
\begin{array}{lll}
\ds\overline{\cal J}^{\rm f}_1(s)&=&\ds{1 \over 24}
B\left( {s+3 \over 2}, -{s \over 2} \right) , \\
\ds {\cal J}^{\rm f}_n(s)&=&\ds \int_0^{\infty} dx \,
x^{-s-1} \, {\cal U}^{\rm f}_n(t(x)) .
\end{array}
\label{JsfromUsf}
\eeq
In order to include $\nu=0$ or
close values, we also find the alternative representation
\beq
\begin{array}{lll}
\zeta_{\nu}^{\rm f}(s)&=&\ds
-{1 \over \pi} \left( \nu + {1 \over 2} \right)^2 {1 \over s+1}
-{1 \over \pi} \left( \nu + {1 \over 2} \right)^2 (-1+ \ln 2)
+{1 \over \pi}{\cal R}^{\rm f}_{\nu}(-1) + {\cal O}(s+1) , \\
{\cal R}_{\nu}^{\rm f}(s)&=&\ds
\int_0^{\infty} dx \, x^{-s-1} \, \left\{
\ln\left[
\pi x e^{-2x} \, ( I_{\nu}^2(x) + I_{\nu+1}^2(x) )
\right]
+\left(\nu + {1 \over 2}\right)^2 {1 \over \sqrt{x^2+1} } \right\} ,
\end{array}
\label{alternrf}
\eeq
which is like \req{zetanuID0le}, for bosons.
Note the vanishing of the $s=-1$ pole for $\nu=-1/2$, which happens
because $\zeta_{\nu=-1/2}^{\rm f}(s)=(\pi/2)^{-s}(2^s-1)\zeta_R(s)$
is finite at $s=-1$.

The analogue of expression \req{barzle}
for the fermionic case is
\beq
\begin{array}{ll}
\ds\overline{\zeta^{\rm f}_{\cal M}}(s;\beta)={1 \over \pi a}&\ds\left[
{1 \over 2}\zeta_H(-2,\beta)
+{1 \over 12}\zeta_H(0, \beta)
-\left( {97 \over 20160} + {\pi \over 256} \right) \zeta_H(2,\beta)
+\left(
 {13 \over 20160}
+ {35\pi \over 32768}
\right) \zeta_H(3,\beta)
\right.
\\
&\ds+\sum_{l=0}^{\infty}
S^{\rm f}_4(-1,l+\beta)(l+\beta) \\
&\ds+\left(
-{1 \over 12}
+{\beta \over 4}
+{\pi\over 128}
-\zeta_H(-2,\beta)
-\zeta_H(-1, \beta )
\right)
\left( {1 \over s+1}+\ln a -1 \right) \\
&\ds-{1 \over 24}-{\ln 2 \over 12}
+\beta{\ln 2 \over 4}
-{\psi(\beta) \over 24}
-{\pi\psi(\beta) \over 128}
-\left( 2+\ln 2 \right)\zeta_H(-2, \beta)
-\left( 1+\ln 2 \right)\zeta_H(-1, \beta) \\
&\ds\left.
-\zeta_H'(-2, \beta)
-\zeta_H'(-1, \beta)
-{1 \over 4}\zeta_H'(0,\beta)
+{\cal O}(s+1)
\right] .
\end{array}
\label{barzlef}
\eeq
Actually, with this plus \req{czf3pf} and \req{alternrf} we realize that
\beq
\zeta^{\rm f}_{\cal M}(s ; \alpha)=
{1 \over a} \left[
{1 \over 64}\left( {1 \over s+1} +\ln a \right)
+ q^{\rm f}_{\alpha} +{\cal O}(s+1)
\right] ,
\eeq
i.e. after adding up all the contributions, the residue of the
resulting pole at $s=-1$ is independent of $\alpha$,
like the residues at $s=1$ and $s=2$, and, in consequence,
the dependence of $E_{vac}(\Lambda )$ on
$\alpha$ is given by $-\frac{1}{2a}q^{\rm f}_{\alpha}$
(remember the minus sign associated to Dirac particles, see
for instance ref.\cite{Greiner}),
where $q^{\rm f}_{\alpha}$ is the remaining finite part
of the $\zeta$-function once the pole has been
removed and does ---predictably---
depend on the magnetic field. Numerical values are given in table 
\ref{tabqa}.
\begin{table}[htbp]
\begin{center}
\begin{tabular}{|l|r|}
\hline
$\alpha$&$q^{\rm f}_{\alpha}$ \\ \hline
0 & $-0.00583$ \\
0.1 & $+0.05603$ \\
0.2 & $+0.07735$ \\
0.3 &$ +0.07753$ \\
0.4 & $+0.06656$ \\
0.5 & $+0.05045$ \\
0.6 & $+0.03314$ \\
0.7 & $+0.01733$ \\
0.8 & $+0.00484$ \\
0.9 & $-0.00312$ \\
1 & $-0.00583$ \\ \hline
\end{tabular}
\end{center}
\scapt{Finite part $q_{\alpha}^{\rm f}$ of the fermionic complete zeta
function for varying reduced flux $\alpha$.}
\label{tabqa}
\end{table}
The fact that $q^{\rm f}_{1}=q^{\rm f}_{0}$, comes from the equality
$\zeta^{\rm f}_{\cal M}(s ; 0)=
\zeta^{\rm f}_{\cal M}(s ; 1)$,
which is in the end a consequence of the modified
Bessel function identity $I_{-n}=I_{n}$, $n=1,2,3, \dots$.
The values of $q_{\alpha}^{\rm f}$ are shown in Fig. 3a. For the
sake of clarity we also give a plot for an extended domain of $\alpha$
which illustrates the particular periodicities of the fermionic case
\req{periodferm}, see Fig. 3b.

\section{Ending comments}
A scalar Klein-Gordon field subject
to Dirichlet boundary conditions and under the
influence of an external magnetic field producing a single flux line
has been studied in two- and three-dimensional spaces.
We have obtained a nontrivial effect in the dependence of
the vacuum energy on the flux which would be invisible (within
the order of our approximation) without the presence of a finite-sized bag.
Considering only the
internal field modes in the $D=2$ case, we arrive at the conclusion
that the vacuum energy undergoes a finite variation when the
magnetic flux is changed. It is interesting to note that around
$\alpha = 0$, the vacuum energy decreases when the flux grows.
In this sense, the system would seem to energetically favour
the presence of such fluxons.

In $D=3$
the flux line is diametrically threading a sphere
and we have quite a different resulting picture.
Taking just the internal modes,
we see that it is not true that the divergent pieces are independent
of the magnetic flux. In fact we have seen that the coefficient
giving the logarithmic divergence in $\Lambda $ (or if you prefer,
the residue of the $\zeta$-function at $z=-1$) depends non-analytically
on $\alpha$. This is also the case for the coefficient giving
a $\Lambda^2$ divergence, associated to the residue at $s=1$.
The inclusion of external modes dramatically modifies the situation.
Their associated divergences exactly cancel those from the internal part,
except the one arising from the pole at $s=2$, but this piece
does not contain any dependence on $\alpha $.
At $\alpha=0$, our result agrees with the one found in ref. \cite{BeMi}.
Around this point the system seems to oppose to
the growth of the kind of fluxons we have pictured, in the sense that
some amount of energy must be provided.

Fermions in $D=2$ have also been considered. While the bosonic energy
was periodic in $\alpha$ with period$=1$, 
the fermionic one
is ---not too surprisingly--- periodic with the same period only
on each real semiaxis separately as shown in Fig. 3b, where
$q_{\alpha}^{\rm f}$ is represented.
The divergences of $E_{reg}(\Lambda)$ are independent
of $\alpha$, with the same
transparency in the physical interpretation of the
result as in the $D=2$ bosonic case.
For values in a neighbourhood of $\alpha =0$, the energy is seen to
decrease as the absolute value of the flux grows. So we have that
the fermionic case in $D=2$ shares with the scalar one this 
property.

Refs. \cite{Mi} include a study of the three-dimensional bag
involving gauge (bosonic) and fermionic massless fields without external
flux.
By way of rough comparison with some of
the figures obtained in these works we may evaluate the ratio between
the maximum variation of the vacuum energy for Dirac field
and the same quantity for a complex Klein-Gordon field.
Taking into account that this variation is given by $0.0397 \over a$
for the fermionic case, and $0.0975 \over a$ for the scalar one,
we have that the ratio is $0.41$. In other words, the energy of
a Klein-Gordon field is in this sense more sensitive to 
flux changes.

To finish this work we shall briefly comment how the analysis that we
have performed in this article with models
without a mass term carries over to cases where this term is present.
Of course, had we incorporated a mass term, the result for the
finite contributions to the zeta functions would be different 
and would call for some extra, though feasible,  effort (see
\cite{MBEEKKSL}). The divergent pieces would also change, but
in a way which is quite trivial. In general, the residue of a pole
at a point $s$, would be transformed into itself plus
a linear combination of the residues of the poles at $s+2k$ for
positive integer $k$'s, with coefficients given by even powers
of the mass. In other words, if one finds that in the massless
case divergences are $\alpha $ independent, this property
carries over to the massive case.

\vskip4ex
\noindent{\Large \bf Acknowledgements}
\vskip3ex
I. Brevik, E.N. Bukina, E. Elizalde, A. A. Kvitsinsky and K.A. Milton
are thanked for comments and discussions.
S.L. gratefully acknowledges an FI grant from
{\it Generalitat de Catalunya}.

\vskip5ex
\noindent{\Large \bf Figure Captions}
\vskip3ex
\ni{\bf Fig. 1}. Bosonic zero-point
energy $E_C$ in $D=2$ for
$a=1$  (then $E_C = q_{\alpha}$) as a function of the reduced flux $\alpha$.

\ni{\bf Fig. 2}. Description of $\zeta_{\cal M}(s ; \alpha)$
---for the $D=3$ internal modes only--- at $s=-1$,
as a function of the reduced flux $\alpha$:
{\bf a)} residue $r_{\alpha}$,
{\bf b)} finite part $p_{\alpha}$,
{\bf c)} comparison of $p_{\alpha}$ for internal and external
modes, together with the Casimir energy $E_{C}$
($aE_{C}=
p_{{\alpha} \, {\rm int}}+p_{{\alpha} \, {\rm ext}}$),
in $D=3$, where $p_{{\alpha} \, {\rm int}}$ is the same as in {\bf b)}.

\ni{\bf Fig. 3}.
{\bf a)} Finite part of the fermionic
zero-point energy $E_{C}=-{1 \over 2a}q_{\alpha}^f$ in $D=2$ for
$a=1$
as a function of the reduced flux $\alpha$.
{\bf b)} Same as in {\bf a}; we have simply
enlarged the domain of $\alpha$ considered.

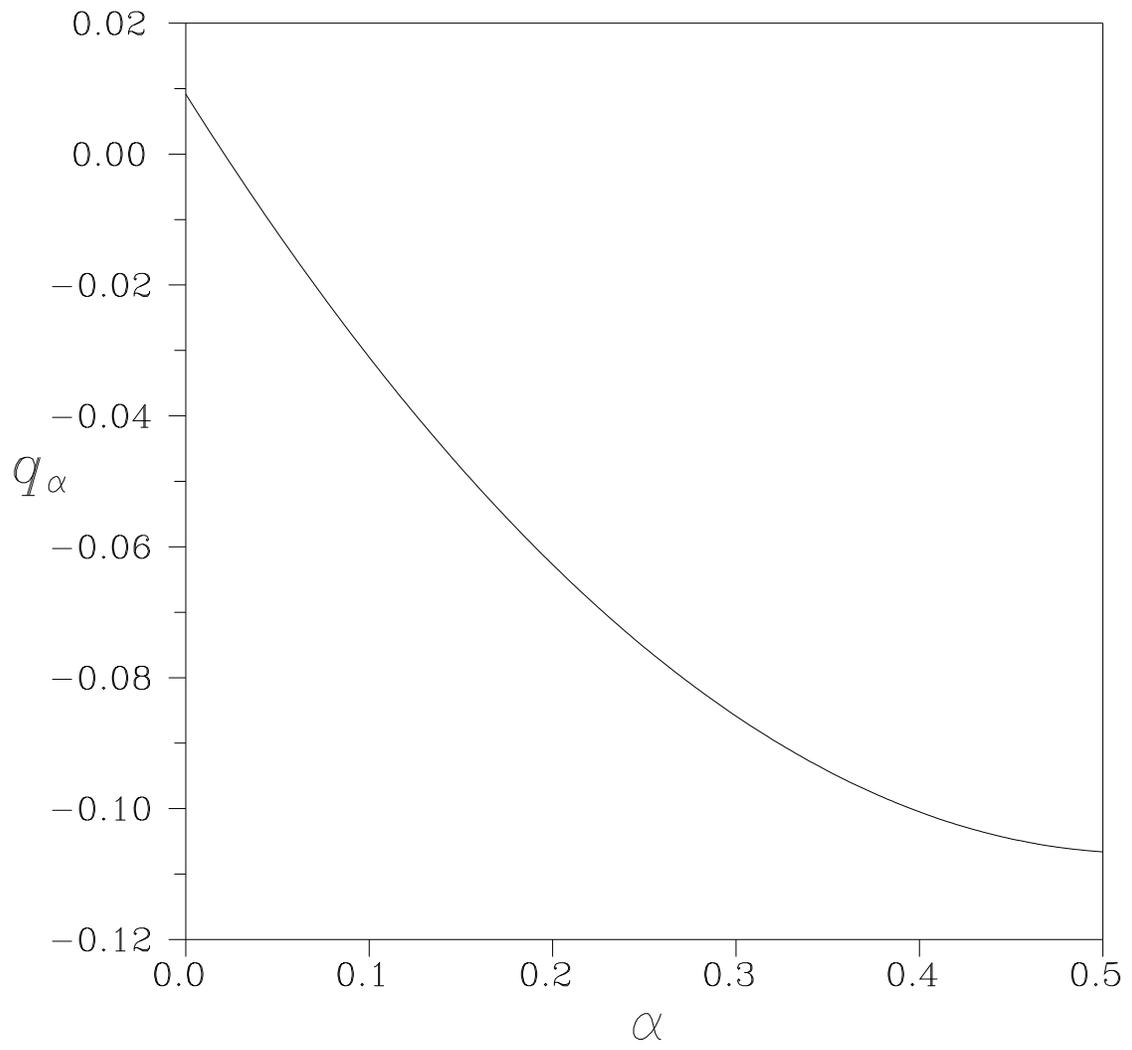
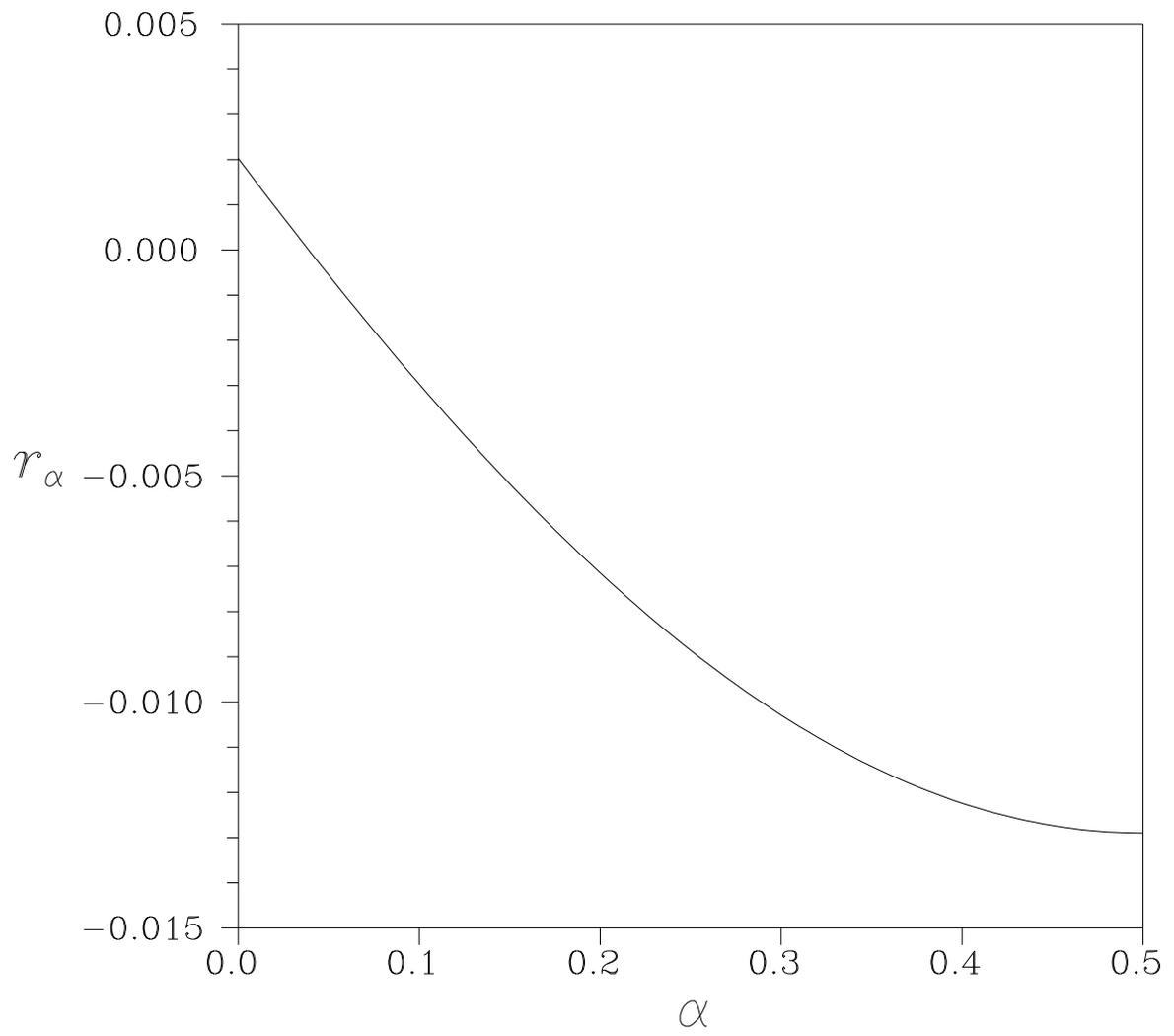
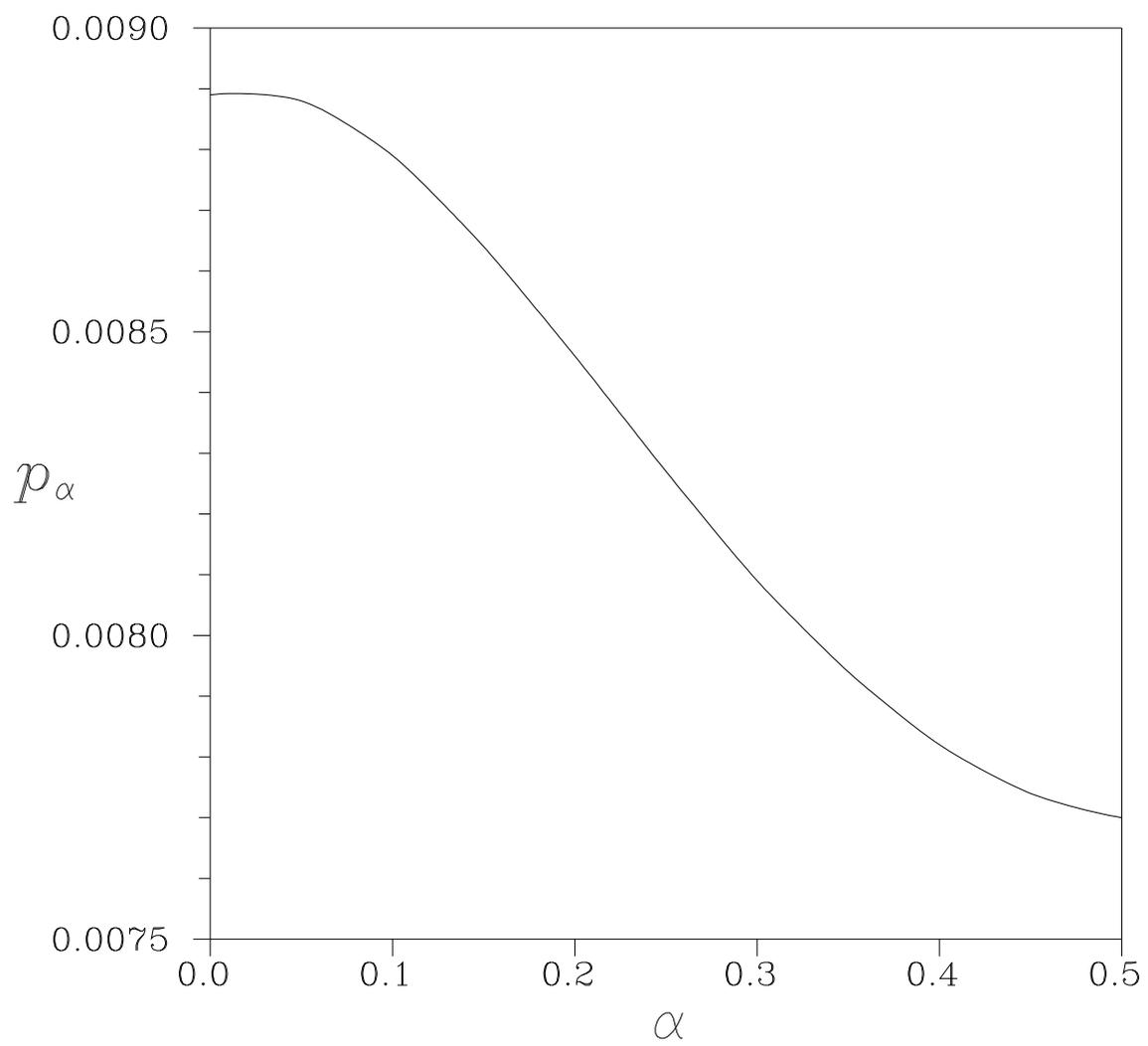
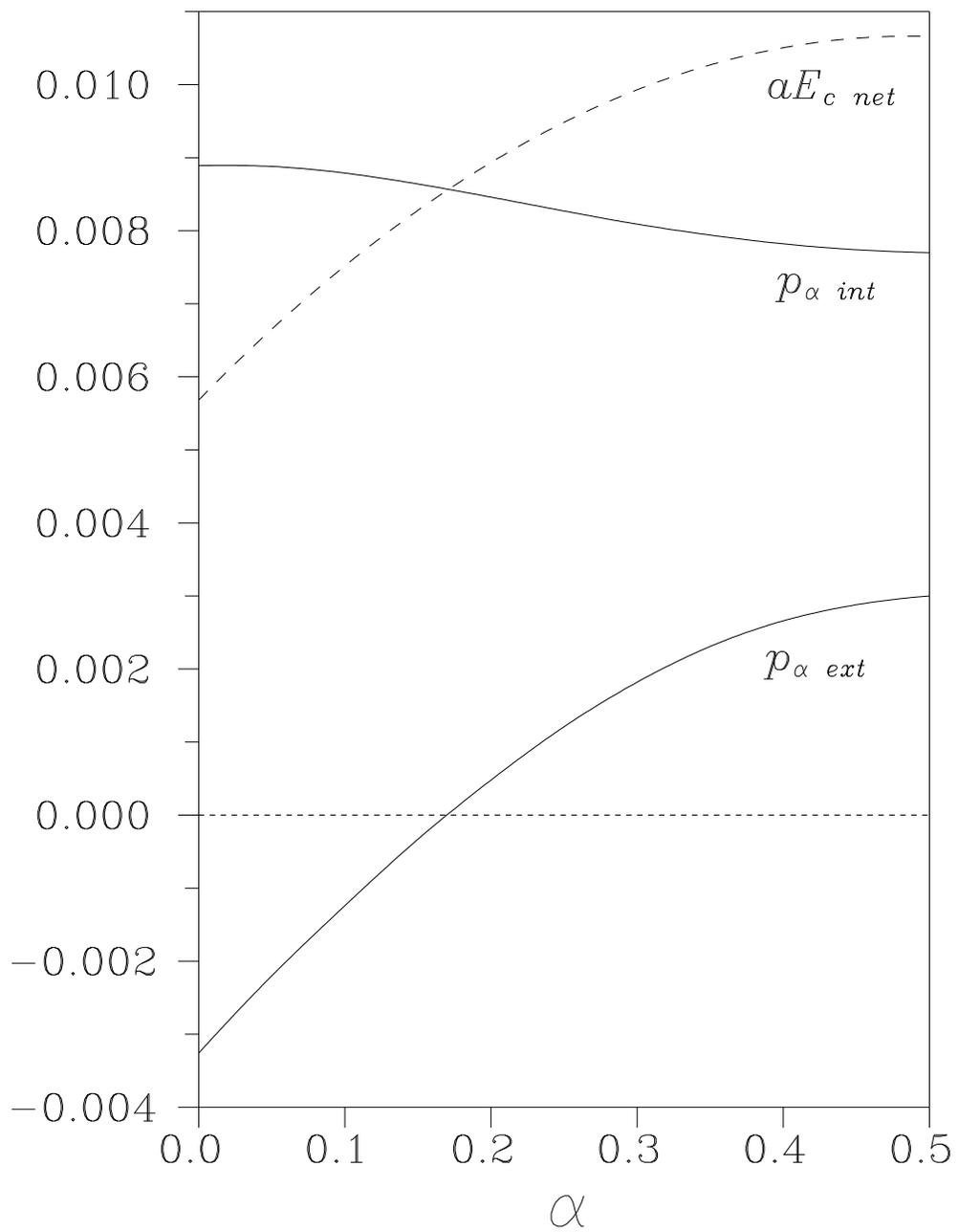
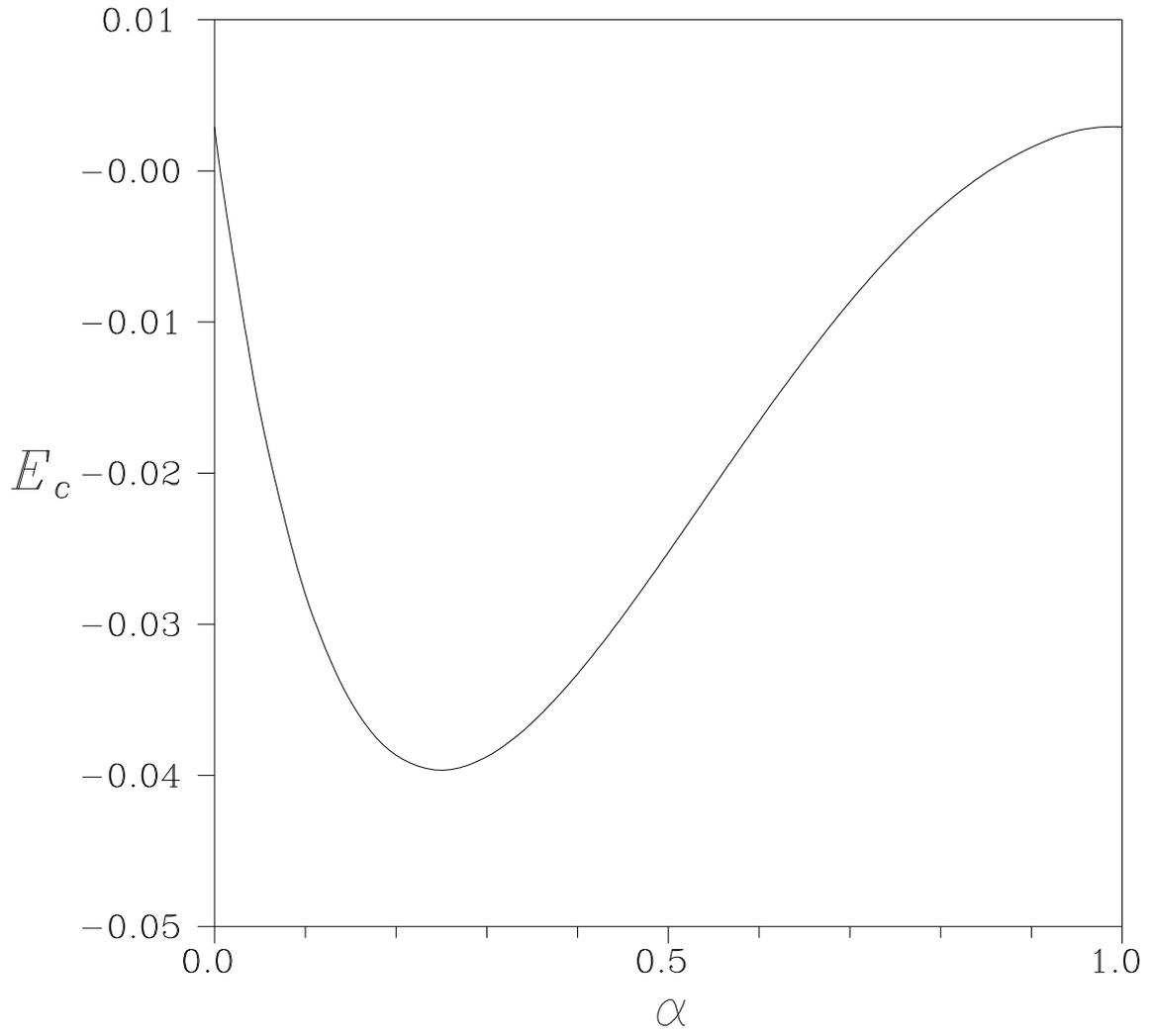
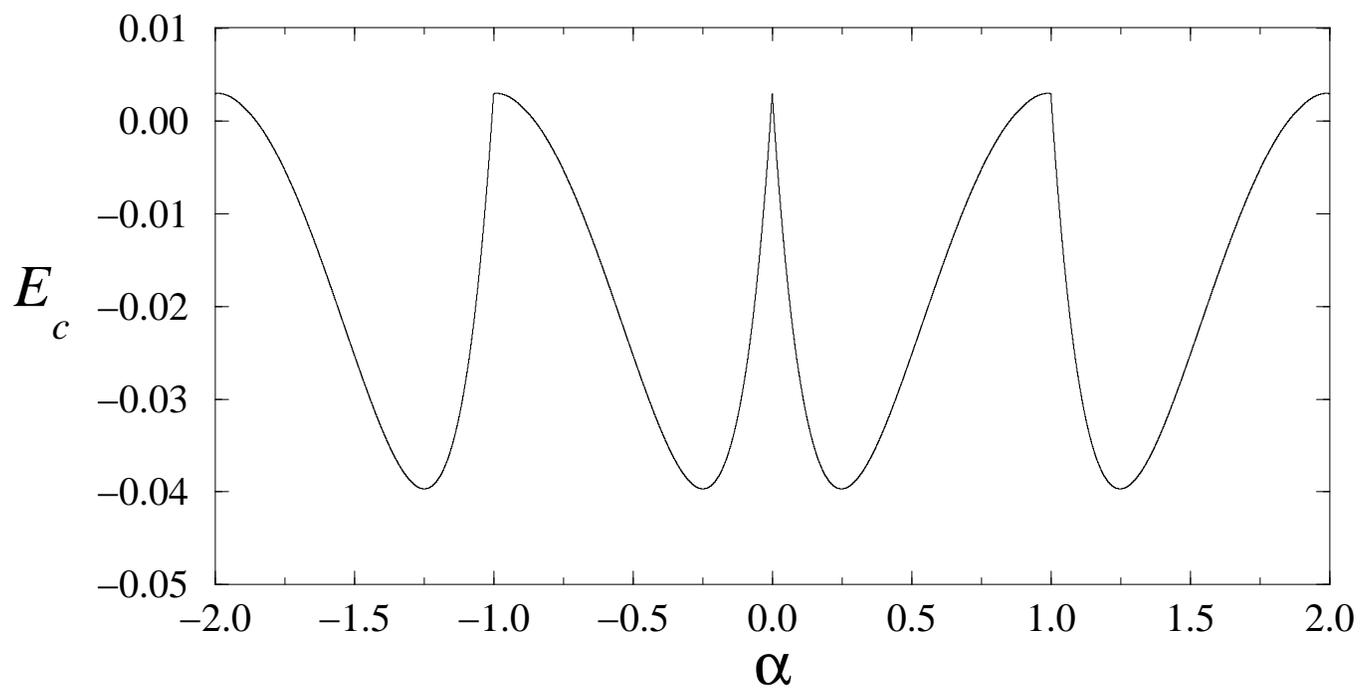
\end{document}